\documentclass[twocolumn,superscriptaddress,showpacs,preprintnumbers,amsmath,amssymb,aps,prd]{revtex4}  
\usepackage{graphicx}
\usepackage{bm}

\def\ltsima{$\; \buildrel < \over \sim \;$}
\def\ltsim{\lower.5ex\hbox{\ltsima}}
\def\gtsima{$\; \buildrel > \over \sim \;$}
\def\gtsim{\lower.5ex\hbox{\gtsima}}

\def\newacronym#1#2#3{\gdef#1{#3 (#2)\gdef#1{#2}}}

\newacronym{\NSF}{NSF}{National Science Foundation}
\newacronym{\NASA}{NASA}{National Aeronautics and Space Administration}
\newacronym{\lisa}{LISA}{the Laser Interferometer Space Antenna}
\newacronym{\ligo}{LIGO}{Laser Interferometer Gravitational-wave Observatory} 
\newacronym{\Caltech}{Caltech}{California Institute of Technology}
\newacronym{\MIT}{MIT}{Massachusetts Institute of Technology}
\newacronym{\sph}{SPH}{smooth particle hydrodynamics}
\newacronym{\tsi}{TSI}{the Terascale Supernova Initiative}
\newacronym{\wmap}{WMAP}{the Wilkinson Microwave Anisotropy Probe}
\newacronym{\decigo}{DECIGO}{the Deci-Hertz Interferometric Gravitational-wave Observatory} 
\newacronym{\cmbr}{CMBR}{cosmic microwave background}
\newacronym{\ibbh}{IBBH}{intermediate binary black hole}
\newacronym{\bdj}{BDJ}{Brans-Dicke-Jordan}
\newacronym{\bbo}{BBO}{Big Bang Observer}
\newacronym{\decigo}{DECIGO}{Deci-Hertz Gravitational-Wave Observatory}

\def\MPR#1{{\it Moving Puncture Recipe}#1 (MPR#1)\gdef\MPR{MPR}}
\def\ahz#1{apparent horizon#1 (AH#1)\gdef\ahz{AH}}
\def\CLA#1{close-limit approximation#1 (CLA#1)\gdef\CLA{CLA}}
\def\NR#1{numerical relativity#1 (NR#1)\gdef\NR{NR}}
\def\pnw#1{post-Newtonian#1 (PN#1)\gdef\pnw{PN}}
\def\qnm#1{quasi-normal mode#1 (QNM#1)\gdef\qnm{QNM}}
\def\isco#1{innermost stable circular orbit#1 (ISCO#1)\gdef\isco{ISCO}}
\def\eos#1{equation of state#1 (EOS#1)\gdef\eos{EOS}}
\def\tov#1{Tolman-Oppenheimer-Volkoff#1 (TOV#1)\gdef\tov{TOV}}
\def\ns#1{neutron star#1 (NS#1)\gdef\ns{NS}}
\def\bbh#1{binary black holes#1 (BBH#1)\gdef\bbh{BBH}}
\def\bhns#1{black hole -- neutron star#1 (BHNS#1)\gdef\bhns{BHNS}}
\def\nsns#1{neutron star -- neutron star#1 (NSNS#1)\gdef\nsns{NSNS}}
\def\emri#1{extreme mass-ratio inspiral#1 (EMRI#1)\gdef\emri{EMRI}}
\def\emrb#1{extreme mass-ratio binaries#1 (EMRB#1)\gdef\emrb{EMRB}} 
\def\grb#1{gamma-ray burst#1 (GRB#1)\gdef\grb{GRB}}
\def\imbh#1{intermediate mass black hole#1 (IMBH#1)\gdef\imbh{IMBH}}
\def\smbh#1{supermassive black hole#1 (SMBH#1)\gdef\smbh{SMBH}}
\def\bh#1{black hole#1 (BH#1)\gdef\bh{BH}}
\def\ulx#1{ultra-luminous x-ray source#1 (ULX#1)\gdef\ulx{ULX}}
\def\lmxbs{low-mass x-ray Binaries (LMXBs)\gdef\lmxbs{LMXBs}\gdef\lmxb{LMXB}} 
\def\lmxb{low-mass x-ray Binary (LMXB)\gdef\lmxbs{LMXBs}\gdef\lmxb{LMXB}} 

\usepackage{color}

\def\amax{$a/M_h \approx 0.823\,$}
\def\lmax{$L/M^2_h \approx 1.176\,$}

\begin{document}

\title{Binary Black Hole Encounters, Gravitational Bursts and Maximum Final Spin}

\author{Matthew C. Washik}
\author{James Healy}
\author{Frank Herrmann}
\author{Ian Hinder}
\author{Deirdre M. Shoemaker} 
\author{Pablo Laguna} 
\affiliation{Center for Gravitational Wave Physics\\
  The Pennsylvania State University, University Park, PA 16802}
\author{Richard A. Matzner}
\affiliation{Center for Relativity and Department of Physics\\
  The University of Texas at Austin, Austin, TX 78712} 

\begin{abstract}
  The spin of the final black hole in the coalescence of nonspinning
  black holes is determined by the ``residual'' orbital angular
  momentum of the binary.  This residual momentum consists of the
  orbital angular momentum that the binary is not able to shed in the
  process of merging. We study the angular momentum radiated, the spin
  of the final black hole and the gravitational bursts in a sequence
  of equal mass encounters. The initial orbital configurations range
  from those producing an almost direct infall to others leading to
  numerous orbits before infall, with multiple bursts of radiation
  while merging. Our sequence consists of orbits with fixed impact
  parameter. What varies is the initial linear, or equivalently
  angular, momentum of the black holes. For this sequence, the final
  black hole of mass $M_h$ gets a maximum spin parameter \amax, with
  this maximum occurring for initial orbital angular momentum \lmax.
\end{abstract}

\pacs{04.60.Kz,04.60.Pp,98.80.Qc} 

\maketitle 

A few years ago, after a decades-long period of development,
breakthroughs were made in computational modeling of strong
gravitational fields that now allow numerical relativists to
successfully simulate \bbh{} from inspiral through merger. In general
terms, there are now two computational recipes to follow. One of them
is based on a generalized harmonic formulation of the Einstein
equations \cite{Pretorius:2006tp,2005PhRvL..95l1101P} and uses
\emph{excision}~\cite{Brandt00,Shoemaker2003a} of the \bh{}
singularities. The other recipe, called the moving puncture recipe,
involves a BSSN \cite{Shibata88,Baumgarte98a} formulation,
\emph{punctures} to model \bh{} singularities and a gauge condition
for these punctures to move throughout the computational domain
\cite{Campanelli:2005dd,Baker:2005vv}. Using these recipes, many
studies involving interacting \bh{s} and their generated gravitational
radiation have been carried out, including gravitational
recoil~\cite{Baker:2007gi,Gonzalez:2006md,Herrmann:2007ac,
  Baker:2006vn,Campanelli:2007cga,Pollney:2007ss,Koppitz:2007ev}, spin
hang-up~\cite{Campanelli:2006uy} and matches to \pnw{}
approximations~\cite{Baker:2006ha,Hannam:2007ik}. Most center on
astrophysical implications and connection to future gravitational wave
observations.

\bbh{} simulations also enable studies of strong non-linear phenomena
regardless of traditional gravitational astrophysics consequences. A
recent example is the work in Ref.~\cite{Pretorius:2007jn} on the
self-similar behavior found in the approach to the merger/flyby
threshold of \bbh{s}. Similar merger thresholds in \bbh{} encounters
or scatterings form the context for our work.

We consider orbits in which the \bh{s} initially fly past one another,
but then fall back to orbit and merge. We focus on the gravitational
waveform and the angular momentum radiated from such
encounters. Serendipitously, we find significant astrophysical
implications, both the existence of a maximum in the final \bh{} spin
and of multiple encounter orbits with associated multiple bursts of
gravitational radiation.  Ref.~\cite{Pretorius:2007jn} considered only
the first close encounter or ``whirl,'' and the study did not extend
the evolutions to find possible fall-back orbits such as those here
considered. The work in Ref.~\cite{Pretorius:2007jn} and our work here
have to date been the only studies considering these highly eccentric
orbits; while there have been high-order \pnw{} studies of inspiral,
cases studied so far have described relatively smooth
inspirals~\cite{2006PhRvD..73l4012K}.

All our orbits are parabolic or hyperbolic encounters.  Depending on
the merger, the fraction of angular momentum radiated varies
significantly ($0.05 \ltsim J_{rad}/L \ltsim 0.55$ with $L$ the
initial orbital angular momentum of the binary). This emission of
angular momentum sets an upper limit of \amax for the spin parameter
of the final \bh{;} this maximum occurs when \lmax, with $M_h$ the
mass of the final merged \bh{.}

As in our previous \bbh{}
studies~\cite{Herrmann:2007ac,Herrmann:2007ex,Herrmann:2006ks}, we use
a code based on the BSSN formulation and the moving puncture
recipe. The results here were obtained with a $634^3\,M$ computational
domain consisting of 10 refinement levels, with finest resolution of
$M/52$. We set up nonspinning equal mass \bh{s} using Bowen-York
initial data~\cite{Bowen:1980yu}. The mass of each \bh{} is $M/2$,
computed from $\sqrt{A_{ah}/ 16\pi}$ with $A_{ah}$ their apparent
horizon area. The data have the \bh{s} on the $x$-axis: \bh{$_\pm$} is
located at $\pm 5\,M$ and has linear momentum $\vec{P}_\pm = (\mp
P\,\cos{\theta},\pm P\,\sin{\theta},0)$. We keep the angle constant at
$\theta=26.565^{\circ}=\tan^{-1}(1/2)$; thus the impact parameter is
$\sim 4.47\,M$. The total initial orbital angular momentum is given by
$\vec{L}/M^2 = 10\,(P/M)\,\sin{\theta}\,\hat z$. We obtain a
one-parameter family of initial data by varying the magnitude of the
initial momentum in the range $0.1145 \le P/M \le 0.3093$. At the
lower limit of the momenta, merger occurs within less than half an
orbit of inspiral. We then consider successively higher initial
momentum until we find solutions that will clearly require a very
long, ``infinite'' time to merge.

\begin{figure}
\begin{center}
\includegraphics[width=10.0cm,angle=0]{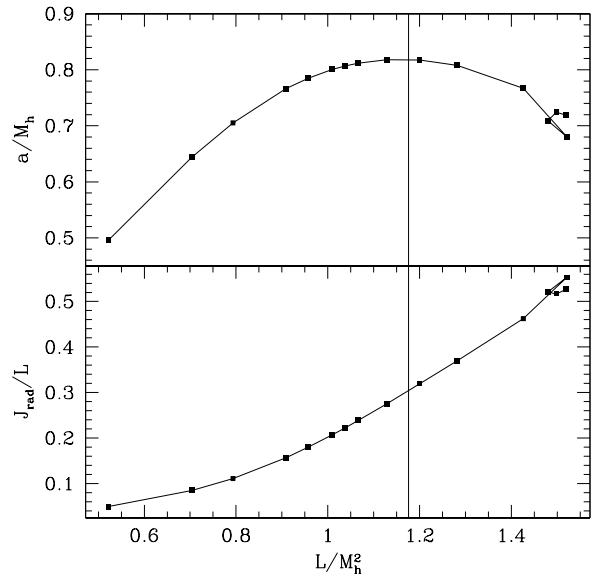}
\end{center}
\caption{Top panel, spin of the final \bh{} $a/M_h$ and, bottom panel,
  angular momentum radiated $J_{rad}/L$ {\it vs} the initial orbital
  angular momentum $L/M_h^2$.}
\label{fig:spin}
\end{figure}

\begin{figure}
\begin{center}
\includegraphics[width=10.0cm,angle=0]{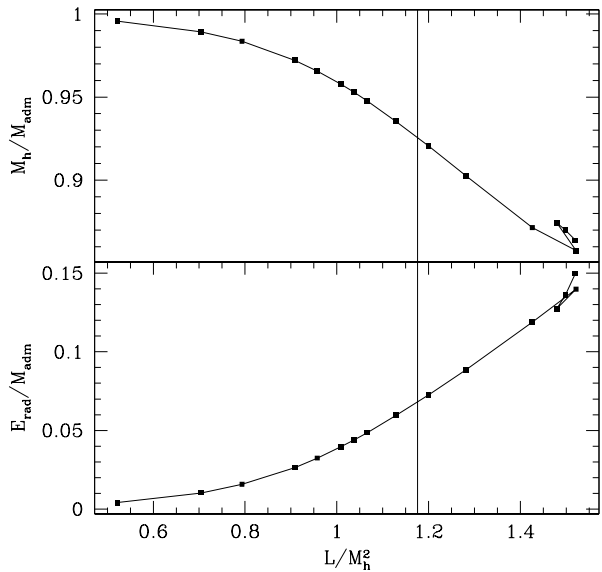}
\end{center}
\caption{Top panel, mass of the final \bh{} $M_h/M_{adm}$ and, bottom
  panel, energy radiated $E_{rad}/M_{adm}$ {\it vs} the initial
  orbital angular momentum $L/M_h^2$.}
\label{fig:ener}
\end{figure}

The results are summarized in Figs.~\ref{fig:spin} and
~\ref{fig:ener}. The top panel in Fig.~\ref{fig:spin} shows the spin
$a/M_h$ of the final \bh{} as a function of the initial orbital
angular momentum $L/M_h^2$. The spin and mass of the final \bh{} were
computed using the apparent horizon
formula~\cite{Campanelli:2006fy,Herrmann:2007ex}. The bottom panel of
Fig.~\ref{fig:spin} displays the fraction of angular momentum radiated
($J_{rad}/L = 1-a\,M_h/L$). Figure~\ref{fig:ener} shows, as a function
of $L/M_h^2$, in the top panel the final mass $M_h/M_{adm}$ relative to
the total ADM mass and in the bottom panel the fraction of energy
radiated $E_{rad}/M_{adm}= 1-M_h/M_{adm}$. The vertical lines in
Figs.~\ref{fig:spin} and ~\ref{fig:ener} denote the value of $L/M_h^2$
where $a/M_h$ is maximum. We have also calculated both the radiated
angular momentum and energy via the Weyl tensor. The results are
consistent with those in Figs.~\ref{fig:spin} and
~\ref{fig:ener}. However, the values obtained form $J_{rad}/L =
1-a\,M_h/L$ and $E_{rad}/M_{adm}= 1-M_h/M_{adm}$ are more accurate
because they are not as susceptible to resolution effects as those
derived from wave extraction. We have carried out simulations at
resolutions of $M/45$, $M/48$, $M/52$ and $M/64$ for ten
representative cases in Figs.~\ref{fig:spin} and~\ref{fig:ener} to
check convergence and make error estimates. We found that the results
are consistent with the 4th-order accuracy of our code and that the
errors in the quantities displayed in these figures are not larger
than 3\%.

We have selected six encounters that are representative of the
different behaviors in our series. These six cases are $L/M^2 =$
0.512, 1.208, 1.352, 1.376, 1.382 and 1.387 or equivalently $L/M_h^2
=$ 0.521, 1.282, 1.522, 1.480, 1.498 and 1.554. We will refer to them
as encounters Ea, Eb, Ec, Ed, Ee and Ef, respectively. Cases Ed, Ee
and Ef correspond to the last three points in Figs.~\ref{fig:spin} and
\ref{fig:ener}. 

\begin{figure}
\begin{center}
\includegraphics[width=10.0cm,angle=0]{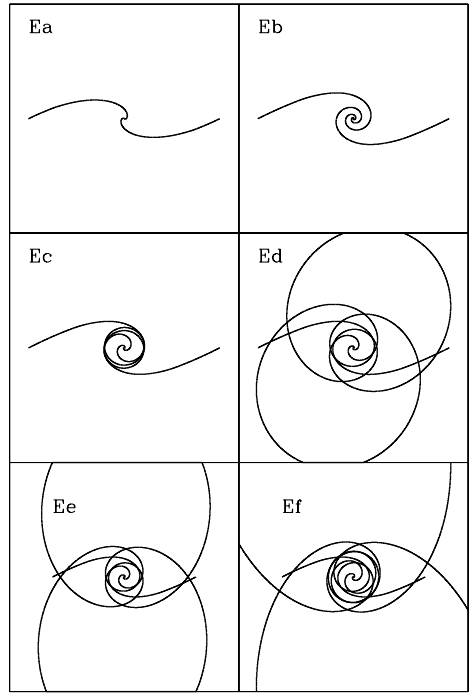}
\end{center}
\caption{\bh{} tracks of the encounters. The coordinate dimensions of
  the top four panels are $12\,M \times 12\,M$ and $16\,M \times
  16\,M$ for the 2 bottom panels.}
\label{fig:path}
\end{figure}

\begin{figure}
\begin{center}
\includegraphics[width=10.0cm,angle=0]{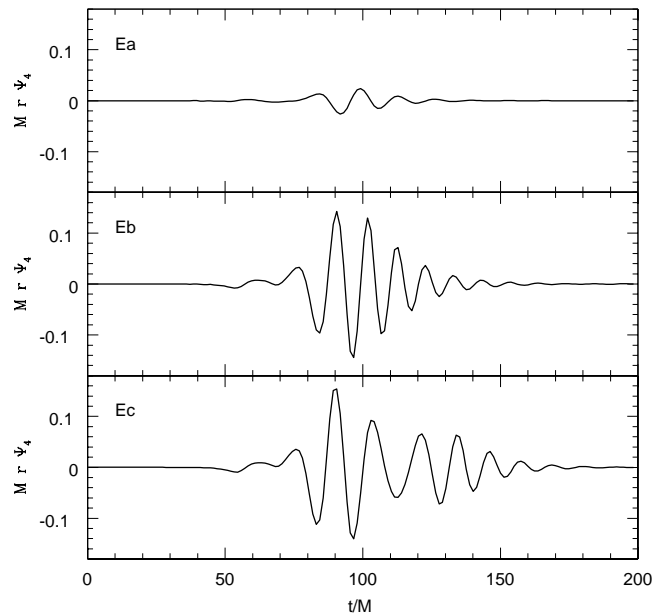}
\end{center}
\caption{Waveforms for the Ea, Eb and Ec encounters.}
\label{fig:gw1}
\end{figure}

For $L/M_h^2 \ltsim 0.8$, the radiated angular momentum is $J_{rad}/L
\ltsim 0.15$, so the final \bh{} has $a/M_h$ close to $L/M_h^2$.  The
evolution is rather simple in these cases: immediate merger, with
minimal inspiral. For instance, in case Ea (Fig.~\ref{fig:path}),
$L/M_h^2 = 0.521$, and $J_{rad}/L = 0.05$; thus most of the angular
momentum goes into the final \bh{,} $a/M_h = 0.496$.
Fig.~\ref{fig:gw1}.Ea shows the corresponding radiated gravitational
wave ($M\,r\, ${\tt Re}$\Psi_4^{2,2}$). All waveforms were extracted
at radius $50\,M$.

As the initial angular momentum increases, the radiated angular
momentum also increases, suppressing and limiting the spin of the
final \bh{.} Eventually for large enough initial angular momentum, so
much angular momentum is radiated that, as seen in
Fig.~\ref{fig:spin}, the final spin reaches a maximum of \amax at
\lmax. Fig.~\ref{fig:path}.Eb shows the tracks of the \bh{s} in the
neighborhood of this maximum. Fig.~\ref{fig:gw1}.Eb shows the
corresponding radiated waveform.  For even larger initial angular
momentum, the spin of the final \bh{} actually decreases for
increasing $L/M_h^2 $. The reason is that the merger is not only
preceded by several hang-up
orbits~\cite{Campanelli:2006uy,Pretorius:2007jn}, but also the merger
yields a highly distorted \bh{} that radiates copiously as it settles
down. Case Ec with $a/M_h \approx0.68$ and $L/M_h^2 \approx 1.522$
represents this situation in which almost 50\% of the initial angular
momentum is radiated (see path in Fig.~\ref{fig:path}.Ec and radiated
waveform in Fig.~\ref{fig:gw1}.Ec).

\begin{figure}
\begin{center}
\includegraphics[width=10.0cm,angle=0]{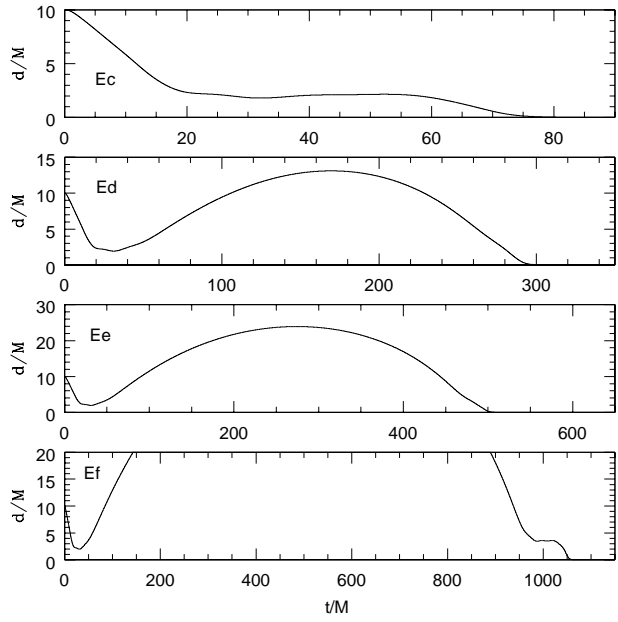}
\end{center}
\caption{\bbh{} coordinate separation for the Ec, Ed, Ee and Ef encounters.}
\label{fig:sepa}
\end{figure}

A persistent feature of the mergers with $L/M_h^2 \ltsim 1.3$ is that
the separation between the \bh{s} (the coordinate distance between the
punctures) decreases monotonically with time (monotonic
inspiral). Comparing cases Ea, Eb and Ec in Fig.~\ref{fig:gw1}, we see
general qualitative agreement: inspiral-generated gravitational waves
with frequency and amplitude increasing in time, followed by
essentially fixed-frequency ringdown waves. There is, however, a hint
of disappearance of the monotonic spiral in case Ec. The amplitude of
the gravitational radiation has a ``shoulder'' at about time $\sim
110\,M$. For a period of time equal to two wave oscillations, the
decline of the amplitude ceases and then recommences. The relative
orbital separation as a function of time (Fig.~\ref{fig:sepa}.Ec)
clearly shows there is a plateau in the separation centered at time
$\sim 50\,M$, which is absent for cases Ea or Eb. For a brief period
of time there is a closely circular phase in which the \bh{s} ``want''
to fly apart, but just manage to stay at roughly constant separation.

\begin{figure}
\begin{center}
\includegraphics[width=10.0cm,angle=0]{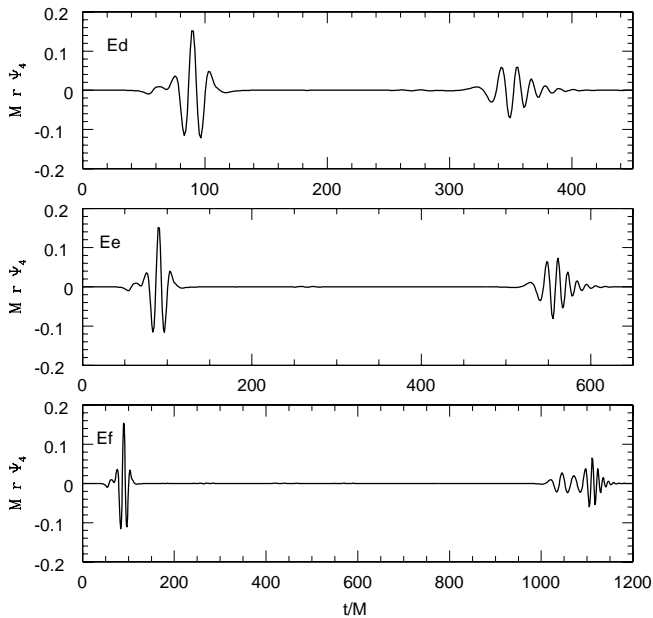}
\end{center}
\caption{Waveforms for the Ed, Ee and Ef encounters.}
\label{fig:gw2}
\end{figure}

The last three points in Figs.~\ref{fig:spin} and~\ref{fig:ener} are
the cases labeled Ed, Ee and Ef. They describe orbits without
immediate merger but ``escape'' and recapture; they all show initial
approaches followed by increasing mid-evolution separations of
$14\,M$, $25\,M$ and $42\,M$ before the final merger (see
Figs.~\ref{fig:path} and~\ref{fig:sepa}). Because the interaction
involves two close approaches, there are two bursts of gravitational
radiation, one from the first flyby~\cite{Yunes:2007zp} and the other
from the merger (see Fig.~\ref{fig:gw2}). We are currently
investigating astrophysical implications of detections of these
multiple gravitational bursts and hangups in globular
clusters~\cite{Kocsis:2006hq}.

For the Ef case, there is an approximate hangup with separation $\sim
4 - 5\,M$ around time $\sim 950\,M$ similar to the shoulder seen in
Fig.~\ref{fig:sepa}.Ec around time $\sim 50\,M$. This structure shows
up in the waveform for this Ef case; we actually see a (lower
amplitude) precursor to the radiation burst associated with the
merger, a hint that orbits with many repeated bounces are
possible. For even slightly ($0.1\%$) greater initial angular momentum
than case Ef, the \bh{s} complete approximately one loop and then
escape. This is a possible indication of chaotic behavior (exponential
dependence on initial conditions, c.f. Ref.~\cite{Pretorius:2007jn}).
Repeated bounce orbits would have to be found with initial angular
momentum very slightly above that which resulted in
Fig.~\ref{fig:path}.Ef. As with all critical phenomena, the problem
becomes one of careful tuning of the parameters.  Note that these
interactions of {\it nonspinning} \bh{s} produce chaotic orbital
dynamics, in contrast to the chaos found in {\it spin}
evolutions~\cite{2000PhRvL..84.3515L,2005PhRvD..71b4027H}.


One of the main conclusions of our work is that {\it there is an upper
  limit on the Kerr parameter} for the final merged \bh{} from
nonspinning \bh{} merger. For our sequence this maximum is \amax.  We
can understand this observation by examining the timing of the
formation of the final \bh{} and the radiation from the merger. It
appears that the merger occurs through an intermediate excited state
which is essentially a highly distorted \bh{.} We say ``essentially''
because a substantial amount of angular momentum is also radiated in
the plunge immediately before the apparent horizon forms. This is
consistent with {\it close limit} \bbh{}
calculations~\cite{Price:1994pm} that show merging \bh{s} behaving
like a perturbed \bh{,} even before a common apparent horizon forms,
so long as the merging \bh{s} are inside the peak of the effective
potential of what will be the final \bh{.} This intermediate state
emits the largest part of the radiated energy and angular
momentum. Because this mechanism is universal (excitation of such a
state is inevitable, and it will inevitably radiate), it suggests that
{\it no} merger of equal mass (or presumably, roughly equal mass)
\bh{s} can lead to a final \bh{} with maximal spin parameter $a/M_h
\approx 1$.  This result does not directly affect spin up by accretion
since mass accretion will not excite the low $l$ modes that strongly
radiate angular momentum.  Thus typical gas accretion can in principle
lead to final spins much closer to the limit $a/M_h = 1$.

Work was supported by NSF grants PHY-0653443 to DS, PHY-0653303,
PHY-0555436 and PHY-0114375 to PL and PHY-0354842 and NASA grant NNG
04GL37G to RAM.  Computations under allocation TG-PHY060013N, and at
the Texas Advanced Computation Center, University of Texas at Austin.
We thank M.~Ansorg, T.~Bode, A.~Knapp and E.~Schnetter for
contributions to our computational infrastructure.

\end{document}